\documentclass[12pt,preprint]{aastex}

\slugcomment{submitted version}

\shorttitle{High-resolution EUV Comet Observations with CHIPS}
\shortauthors{Hurwitz et al.}

\begin{document}


\title{A Search for EUV Emission from Comets 
with the Cosmic Hot Interstellar Plasma Spectrometer (CHIPS)}


\author{T. P. Sasseen}
\affil{Department of Physics, UC Santa Barbara, Santa Barbara, CA, 93106}
\email{tims@ssl.berkeley.edu}

\author{M. Hurwitz}
\affil{Space Sciences Laboratory, University of California,
    Berkeley, CA 94720}

\author{C.M. Lisse}
\affil{Johns Hopkins University Applied Physics Laboratory, 
    Laurel, MD, 20723}

\author{V. Kharchenko}
\affil{Harvard-Smithsonian Center for Astrophysics, 
    Cambridge, MA, 02138}

\author{D. Christian}
\affil{Queens University, Department of Pure \& Applied Physics, Belfast, U. K.}

\author{S. J. Wolk}
\affil{Harvard-Smithsonian Center for Astrophysics, 
    Cambridge, MA, 02138}

\author{M. M. Sirk}
\affil{Space Sciences Laboratory, University of California,
    Berkeley, CA 94720}

\author{A. Dalgarno}
\affil{Harvard-Smithsonian Center for Astrophysics, 
    Cambridge, MA, 02138}



\begin{abstract}
We have obtained EUV spectra between 90 and 255 \AA\/ of the comets 
C/2002 T7 (LINEAR), C/2001 Q4 (NEAT), and C/2004 Q2 (Machholz) near 
their perihelion passages in
2004 with the Cosmic Hot Interstellar Plasma Spectrometer (CHIPS). 
We obtained contemporaneous data on Comet NEAT Q4 
with the $Chandra$ X-ray Observatory ACIS instrument, marking 
the first simultaneous
EUV and X-ray spectral observations of a comet. 
The total CHIPS/EUV observing times were 
337 ks for Q4, 234 ks for T7, and 483 ks for Machholz
and for both CHIPS and $Chandra$ we calculate we have captured 
all the comet flux in the instrument field of view. 
We set upper limits on solar wind charge exchange emission
lines of O, C, N, Ne and Fe occurring in the spectral bandpass 
of CHIPS.  The spectrum of Q4 obtained with $Chandra$ 
can be reproduced by modeling emission lines of 
C, N O, Mg, Fe, Si, S, and Ne solar wind ions.  
The measured X-ray emission line intensities are consistent with 
our predictions from a solar wind charge exchange model. 
The model predictions for the EUV emission line intensities are determined 
from the intensity ratios of the cascading X-ray and EUV photons 
arising in the charge exchange processes.  They are compatible
with the measured limits on the intensities of the EUV lines.  
For comet Q4, we measured a total X-ray flux of 3.7$\times 10^{-12}$ ergs cm$^{-2}$ s$^{-1}$,
and derive from model predictions a total EUV flux 
of 1.5$\times 10^{-12}$ erg cm$^{-2}$ s$^{-1}$. 
The CHIPS observations occurred predominantly while the satellite 
was on the dayside of Earth.  
For much of the observing time, CHIPS performed observations at smaller solar
angles than it was designed for and EUV emission from the Sun scattered into
the instrument limited the sensitivity of the EUV measurements.

\end{abstract}




\section{Introduction}

Soft X-ray emission has been detected from at least 20 comets
since 1996.  After the consideration and rejection of several
possible emission mechanisms, a consensus has been reached that 
the primary mechanism is charge exchange collisions between 
highly charged solar wind minor ions and neutral atoms and molecules 
of the cometary atmospheres, as suggested by \citet{Cravens97}.  
The observations have been reviewed by 
\citet{Cravens02}, \citet{Kras04a} and \citet{Lisse04}. 
The solar wind charge exchange (SWCX) process also occurs in the
heliosphere where it contributes to
the soft X-ray background \citep{Cox88, Cravens00, Rob03,
Pepino04, Lalle04} and may be responsible 
for some fraction of the short and long term enhancements
seen at 250 eV \citep{Sno95, Sno04}. 
These recent, interesting observations indicate it
is important to understand both the microphysics of the charge-exchange
process and the phenomenology of where, when and how this process
occurs within the solar system.  

Theoretical models of the X-ray and EUV emission lines produced in comets
by charge exchange with the solar wind ions have been constructed by
\citet{Weg98}, \citet{schwad00},
\citet{khar00}, \citet{khar01}, \citet{khar03}, 
\citet{Beiers03}, and \citet{Lisse04}. 

Observations of diffuse SWCX emission 
other than from comets have been carried out using
$Chandra$ \citep{War04} and XMM-Newton \citep{Sno04}.  
\citet{War04} observe the optically dark side of the moon and attribute
the X-rays they detect to charge exchange collisions 
between solar wind ions
and neutral hydrogen in the Earth's geocorona.  
\citet{Sno04} ascribe the emission they see as coming from
the Earth's exosphere or the heliosphere.  
In both cases, the emission is characterized by strong variability and
correlation with the solar wind particle flux.  
X-rays generated by SWCX have been detected from the atmospheres
of Mars and Venus \citep{mur96, 
Den02, holm04, fok04, Gun04}.

In summary, the highly-ionized, heavy ions that stream out in the 
solar wind are a potential source of soft X-ray and EUV emission from
within the solar system.  Charge exchange emission may occur
involving neutral atoms and molecules from comets 
entering the inner solar system, 
planetary exospheres, and neutral atoms in the inter-planetary medium.   
A highly-ionized solar wind ion may give rise to several photons 
as the ion captures electrons sequentially in its flight from the
Sun.  A detailed ionization and density map of the heavy solar wind
ions as a function of time and position is required 
for a complete description of the X-ray and EUV emissions 
in the solar system. 
Measurements of SWCX emission from comets (and other measurements) 
can provide a useful probe of the density and ion 
composition of the solar wind. 

Many SWCX lines are predicted to appear at 
EUV energies below the low-energy limit of $Chandra$ and XMM-Newton 
(Kharchenko \& Dalgarno 2000, 2001, Kharchenko et al. 2003).  
Some of these EUV lines arise from ions not conclusively 
identified from emission at higher energies. 
These lines may be detected with CHIPS,  
a spectrometer sensitive in the EUV with resolution of several hundred.  
The lines appear to be present in the ROSAT PSPC 
spectrum of C/1990 K1 (Levy) (\citep{Den97}) which rises 
steeply at energies below 200 eV.
\citet{Kras01} observed comet C/1996 B2 (Hyakutake) 
at low resolution at EUV wavelengths 
and claimed detection of lines from O, C, He and Ne.  
At lower energies, \citet{khar01} predict that emission lines
from the \ion{O}{6} doublet at 1032 and 1038 A should be the brightest 
FUV lines SWCX lines. 
A sensitive search for EUV and FUV  emission lines cometary 
SWCX was conducted with the Far-Ultraviolet Spectroscopic
Explorer (FUSE) observatory \citep{Weaver02, Feldman05}.
\citet{Weaver02} report a
a tentative detection of 1032 \AA\/ emission line from comet C/2000 WM1
(LINEAR).   An alternative interpretation of this feature has been 
suggested by \citet{Feldman05} from in FUSE observations of the 
brighter comet C/2001 Q4 (NEAT); they attribute this feature to 
H$_2$ fluorescent emission line pumped by 
radiation from \ion{N}{3} and \ion{O}{6} ions in the solar corona.  
The non-detection
of the brightest FUV \ion{O}{6} lines from SWCX may be because
the relatively small field of view of FUSE allowed observation of only 
about 6000-7000 km around the cometary nucleus.  
For the Q4 FUSE observations, the peak brightness of the 
charge-exchange EUV and X-ray emissions 
was shifted by more than this amount from the comet center, 
hence the FUSE aperture may have missed it \citep{Feldman05}.  

The favorable perigee passages of the bright 
Comets C/2001 Q4 NEAT and C/2002 T7 LINEAR in May 2004 and Machholz in 
January 2005 provided an excellent opportunity to use the unique 
low energy spectroscopic capabilities of CHIPS to study SWCX in 
comets at EUV wavelengths.  Due to a combination of high solar activity and 
their close approaches, Comets C/2001 Q4 NEAT and C/2002 T7 LINEAR are the 
brightest X-ray targets available since Comet C/Hyakutake in 1996 and 
Comet 153P/2002 C1 (Ikeya-Zhang) 
in 2002.  In this paper, we discuss the observational data 
from CHIPS and $Chandra$, describe the procedures by which 
we measure the line strengths and discuss the 
implications for the cometary charge exchange process.  

\section{Observations and Data Reduction} \label{obs}

\subsection{Instrumental}

The CHIPS orbit is approximately circular at 600 km and an inclination of 
94 degrees, determined by the primary payload (ICESAT) of 
the rocket that launched the instrument.  
CHIPS is sensitive to the EUV band of 90 - 260 \AA\/, or 47 - 138 eV.   
With all six slits open, the spectrograph field of view is roughly
rectangular at 5 by 25 degrees. 
Any object smaller than a few degrees presents a "point
source" to CHIPS, and will be visible only through two of the
slits at a time.  The observations presented here were collected in wide-slit mode, 
with a peak resolution for diffuse light of about 3.4 \AA\/ FWHM,
reaching about 4 \AA\/ at the edges of the bandpass.   

Light from the comet enters the CHIPS spectrograph through two slits, each 1
mm by 7 cm.  The light diverges onto an 8 cm $\times $ 6 cm grazing incidence
cylindrical diffraction grating with central groove density of 1800 lines
mm$^{-1}$.  The groove density is varied across the grating surface to
provide nearly stigmatic focusing in the dispersion direction near the
center of the spectral band.  Thus, the spectral resolution near the center
of the band is limited by the slit width.  At the ends of the band, the
combined effects of optical aberration and defocusing 
contribute to the net spectral line width.  For
sources that are significantly smaller than the field-of-view 
of a slit, the resolution is
independent of the angular size of the comet. 

The grating rulings and baffling reject most non-EUV light and 
thin-film filters placed close to the detector also attenuate out-of-band 
and scattered light.  A diffuse emission line spans the full available 
detector height, typically passing through two different filter materials: 
aluminum, zirconium, or polyimide/boron.   Measurements of an emission line 
that has the appropriate ratio of line flux from the two different filters 
spanned by the line encourages the belief that an actual 
spectral line has been detected and not some detector artifact. 

The CHIPS in-orbit calibration is discussed in more detail in 
\citet{hurwitz05} and additional details of the science instrument 
and the CHIPSAT satellite are presented in 
\citet{hurwitz05}, \citet{hurwitz03}, 
\citet{janicik03}, \citet{marckwordt03}, \citet{sirk03}, 
and \citet{sholl03}. 

\subsection{Comet Observations and Data Reduction}
A summary of the CHIPS observations of all three comets is presented
in Table 1. We coadd individual comet observations in different subsets 
as described below. 
Individual pointings are of $\sim$20 minute duration and 
owing to the orbital geometry, nearly all observations took place
while the satellite was on the day side of Earth.  
The CHIPS mission has a nominal constraint of 72 degrees for the minimum 
boresight-to-Sun angle permitted during observations, both for spacecraft 
health reasons and to minimize the amount of scattered solar radiation entering 
the spectrometer.  However, observing NEAT Q4 and LINEAR T7
near their time of maximum optical brightness required observations
that violated this Sun-angle constraint and required the development and
use of a new observing mode to accommodate the smaller solar angles
as well as observe a target moving in the sky. 
The two most important differences for this mode are a slightly reduced 
pointing accuracy and stability since the main Sun sensor 
did not have the Sun in the field of view and a possible increase
in the amount of scattered solar EUV radiation present in the 
comet spectra.  

Initial data reduction was carried out using the pipeline that is
described in \citet{hurwitz05}.  We apply several filters to the 
raw data in producing the reduced spectra.  Data are excluded when the overall
detector count-rate exceeds 80 events/sec, indicative of high charged-particle
backgrounds.  Data are also excluded when the detector high voltage is reduced
(a normal, temporary response to transient high-background periods).
A broad pulse-height filter is applied by the flight software.
Approximately 40\% of the telemetered events are then excluded
by a low-pass pulse-height threshold in ground software, selectively
reducing background because the low amplitude events are overwhelmingly
triggered by charged particles rather than by photons.

Periodically pulsed "stimulation pins" outside the active field of view are
used to register the event X,Y coordinates to a common frame,
thereby correcting for thermal drifts in the plate scale or
axis zero-points, and to determine the effective observing
time of the observation.  Distortion is corrected using a preflight pinhole
grid map.  Small regions of known detector "hot spots" are
excluded, the event coordinates are rotated so that the new X axis
closely corresponds to the spectral dispersion direction, and
the spectra in each filter-half are summed over the active
detector height.

Once the final spectra (one each from the upper and lower detector
halves) have been summed, these will typically contain counts from
several background sources in addition to any signal from a comet.  
The primary backgrounds include particle events that are generally
uniform across the detector, geocoronal He I 584 \AA\/ photons that
are scattered by the gratings and penetrate the Al filters, and
for these comet observations, some EUV photons from the Sun that
are multiply-scattered from some structure on the satellite into the 
spectrometer.  These backgrounds must be properly accounted for or subtracted
before we can assess the amount of radiation from the comets that may
be present.  

\begin{figure}
\includegraphics[angle=90, scale=0.60]{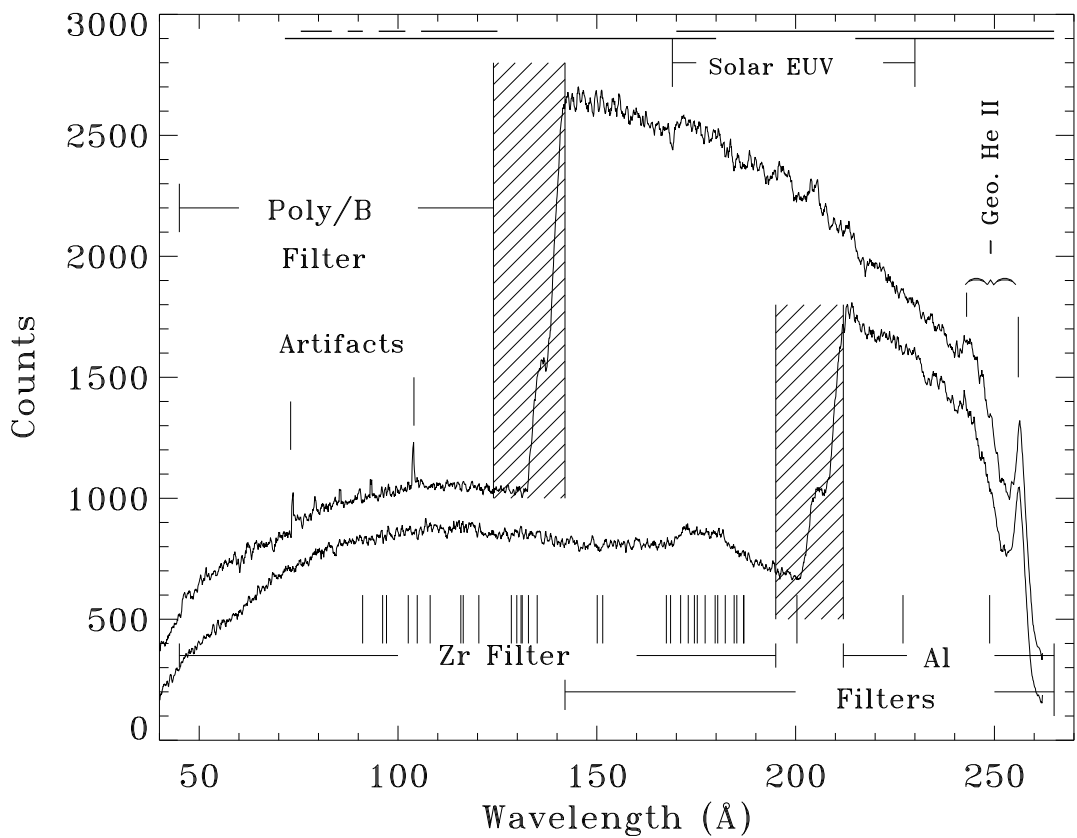}
\caption{Complete CHIPS spectrum after pipeline processing
using all data from C/2001 Q4 (NEAT) on the two detector halves, shown in counts. 
The lower spectrum has been displaced by 200 counts for clarity.  
The filter bandpasses and edges, location of known solar and geocoronal lines, 
and detector/light artifacts that show up in all CHIPS spectra are marked.  
Vertical marks show the location of charge exchange lines predicted
in this band from comets.   Our flux detection limits for a 
particular line use one or both halves (indicated by lines at the top
of the figure, P/Al top, Zr/Al bottom) to optimize the signal-to-noise at 
that wavelength. 
\label{Q4summed}}
\end{figure}

We show in Fig.~\ref{Q4summed}  the summed spectra from all the 
individual NEAT Q4 
observations, totaling 329495 s of useful data.   The upper curve
is the Polyboron/Al side of the detector, while the lower trace
is the Zr/Al side of the detector, lowered by 200 counts.  
The wavelength ranges of the various filters are indicated, as 
are the positions of the filter-supporting bars for each half.  
The continuum seen in the Poly/B and Zr filters is primarily from
particle events, while the Al filters show a significant additional
smooth background from scattered geocoronal He I 584 \AA\/ photons. 
The effective area of the spectrometer as a function of wavelength
is shown in \citet{sirk03}.  We add together signal from one or
both halves of the detector at a particular wavelength based on
the filter transmission curves to maximize signal-to-noise at
that wavelength or to exclude artifacts.  The ranges included from each 
filter/half are indicated in Fig.~\ref{Q4summed} . 

To determine the continuum against which to search
for emission lines, we use a very deep flight charged-particle spectrum,
scaled to match the local counts in the spectrum under analysis.  
The deep charged-particle spectrum generally contains many more events
per wavelength interval than does the spectrum under analysis,
and thus contributes relatively little to the shot noise
in the line flux.  Differences in the technique
by which the scale factor is determined -- for example, heavy smoothing
versus fitting low-order polynomials -- result in changes of $\sim$10\% 
to the line flux results.
To determine the absolute "flatness" of the detector
response, we analyzed a pre-flight photon 
flat field that was histogrammed like the flight spectra.
Compared to a smooth polynomial, the flat field spectrum
showed a pixel-to-pixel variance only slightly greater than
expected for shot noise, where "pixel" refers to the width of
the narrowest spectral features.  The excess   
one-sigma variance above shot noise was about 0.3\% of the total signal.  
This factor is negligible in all the limits reported here,
where shot-noise from charged particles detected is the dominant 
uncertainty.  

We measure the line flux of an individual line by adding all
the counts that occur within one spectral resolution element 
of the line rest wavelength,
and subtracting the counts summed from the scaled background
spectrum over the same limits. 

The wavelength scale is based on preflight measurements, offset by a 
(constant) $\sim$1 \AA\/
determined from the bright geocoronal \ion{He}{2} 256.3 \AA\/
feature, which is present in essentially all the observations.  
Temporal variations in the measured centroid of the \ion{He}{2} feature show a 
dispersion of about 0.2 \AA\/.  
Periodic observations of the moon, which provide a weakly reflected solar 
spectrum
that includes lines of Fe {\sc IX} through {\sc XII}, confirm that the 
adopted
wavelength scale provides a good fit near the center of the spectral
band, and support the view that the relative throughput of the filter panels
is as measured pre-flight.  The measured fluxes of both the \ion{He}{2} feature
and the lunar spectrum are in good agreement with pre-flight expectations,
suggesting that the instrument throughput has not declined since laboratory
calibration.  The pre-flight laboratory throughput calibration, tied
to an NIST-calibrated diode, both at a component level and an 
end-to-end manner (which agree to within $\sim$10\%)
provide the absolute flux calibration.  

\subsection{Pointing Stability}

The CHIPSat attitude control system (ACS) is described in 
\citet{janicik03}.   The ACS uses inputs from several sensors,
include coarse and medium Sun sensors, magnetic sensors, a lunar sensor, and 
reaction wheel rotation sensors to calculate the spacecraft attitude. 
During most of the observations reported here, CHIPS 
was unable to use its most accurate source of 
attitude information, the spacecraft's medium Sun
sensor.  Instead, the adopted observing procedure 
was to orient the spacecraft such that the Sun was 
in the Sun sensor for the fraction of each orbit when a comet
was not being observed, then slew to the comet during the
period the comet was visible.   When the Sun sensor is used, 
the typical pointing accuracy of the spacecraft is 
within $\sim$0.5 degrees of the requested target \citet{janicik03},
and centered to within 1/10th of the spectrometer 
field-of-view. 
For relatively short periods even with slews, such as the length of a 
typical 25 minute comet observation, the spacecraft reaction wheels 
provide pointing orientation and knowledge with negligible degradation. 
This is confirmed by the very small attitude corrections that 
followed reaquisition of the Sun by the medium Sun sensor. 

In the CHIPS observations of the comets, we conclude that 
essentially all of the EUV-emitting region of the comet is contained 
within the CHIPS entrance apertures during the observations. 
The field of view of the two central slits of CHIPS 
is approximately five degrees square.  
Comet positions were generated every 10 minutes during
the period they were observed by CHIPS using 
the Minor Plant and Comet Ephemeris 
Service\footnote{http://cfa-www.harvard.edu/iau/MPEph/MPEph.html}, where
the requested pointing corresponded to
the mid-point in time of each observation. 
Parallax and the motion of the comet during an orbit's pointing 
are negligible compared with the CHIPS field of view. 
Our re-mapped image of Q4 on the $Chandra$ ACIS-S detector
shows the X-ray emitting region to be no larger than 
10 $\times$ 10 arcmin.   
Given the attitude information and the size
of the X-ray emitting region, we believe that 
the comets were positioned in the slit entrance aperture when
the spacecraft reported it was pointed at the requested comet 
position and that the emitting region was fully contained 
within the CHIPS field of view.  Hence, for the observations 
reported here, we assume no dilution factor for comet EUV emission.  
The roll angle for observations was chosen such that the CHIPS entrance 
slits were perpendicular to the great circle that contained both
the comet and the Sun. 

\subsection{Background/Noise Sources}

The majority of counts registered in the CHIPS detectors are
from charged particles entering the instrument.  Although some discrimination
against particle events based on pulse height takes place as
described above, the time-varying 
nature of this background makes it difficult to determine the presence of
any continuum emission arising from photons.  
The net effect of the backgrounds and data reduction techniques 
is that CHIPS is primarily sensitive to spectral 
lines, with a limiting sensitivity of a few 
line units or LU [photons cm$^{-2}$ s$^{-1}$ sr$^{-1}$] for the observations
described here. 

CHIPS consistently detects at high significance He II emission arising 
in the Earth's geocorona at 256.3 \AA\/ and to a lesser degree at 243.0 \AA\/. 
In addition, we observe 
radiation that appears to originate in the solar corona and is multiply 
scattered from surfaces on the spacecraft into the instrument. Although this 
background is very faint compared with the interstellar medium emission 
lines from Fe IX - XII that CHIPS was designed to observe, 
we have studied it in depth because it is a signficant background
to these measurements.   
The EUV line emission from the interstellar
medium reported by \citet{hurwitz05} based on one year of 
CHIPS observations are very faint and 
do not pose a significant background for the comet observations. 
A detailed study of the instrument-scattered background shows that 
the lines are never seen in co-added nighttime-only observations; 
the spectrum appears very similar to the full moon spectrum; 
the strength of the lines does not correlate with zenith angle; 
the lines show no ram-angle dependence; 
the lines appear reproducibly in only a relatively small number
of spacecraft orientations relative to the Sun; 
the lines appear more often, but not always, when CHIPS points closer to
the Sun than during normal observations, but there is not a significant
anti-correlation with solar-target angle; 
the line strengths show no correlation with the 
Ultra Low Energy Isotope Spectrometer (ULEIS) \citep{mas98}
measurements of Fe IX, X and XI in the solar wind during the time of 
the comet observations; and 
the line strengths during comet observations do not
correlate with measurements of solar
disk irradiance in the 170 - 230 \AA\/ by
the Solar EUV Experiment on TIMED (SEE) \citep{woods98}. 
We conclude that the source of these background 
lines is faintly-scattered radiation from the solar corona. 
Owing to its appearance preferentially at smaller solar angles, 
this background is therefore of more importance for the CHIPS comet 
observations than during normal sky survey observations, which 
are made at larger solar angles. 

\begin{figure}
\includegraphics[angle=90, scale=0.60]{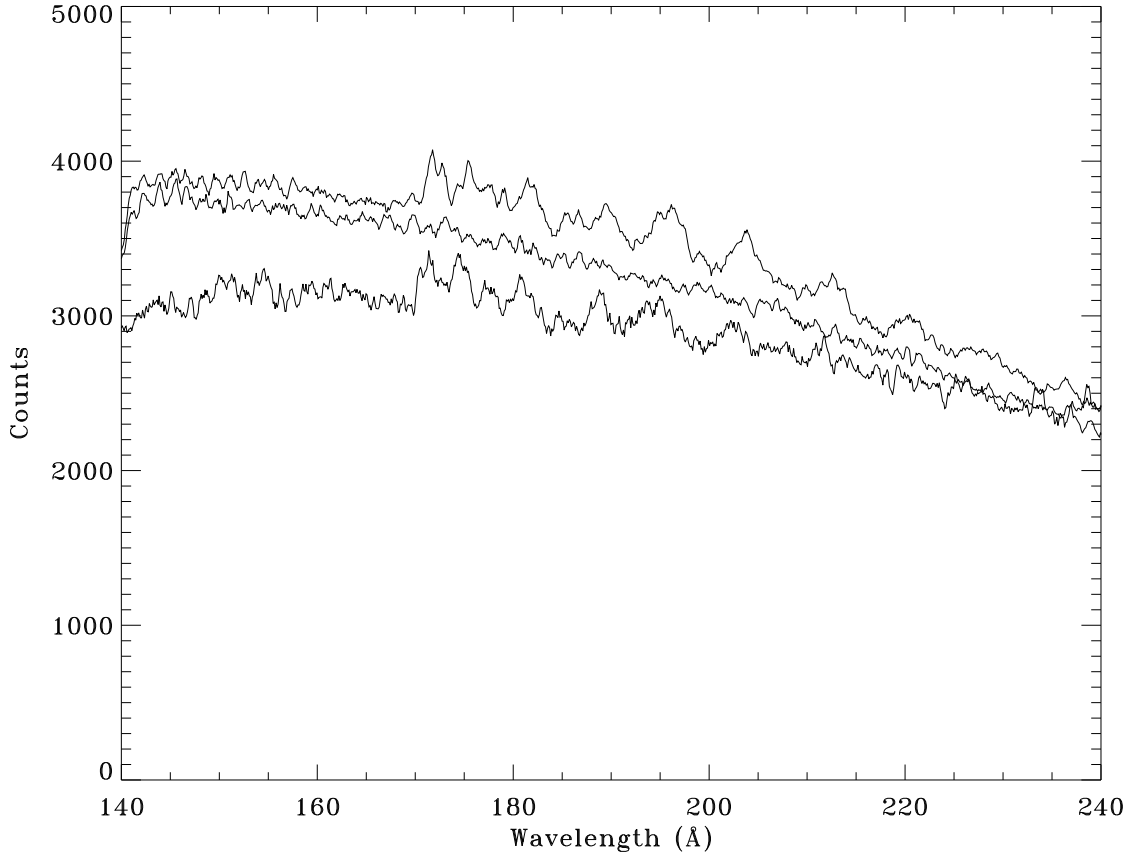}
\caption{CHIPS dayside comet and lunar spectra through Al (polyimide/Boron
detector half) filter and standard pipeline processing. 
The spectra are not divided by the filter response. 
The upper trace shows a set of co-added 
CHIPS observations selected when the dayside background lines are brightest.  
The middle trace shows a set of dayside spectra chosen with
low LFI, showing no solar contamination. 
The lowest trace shows the co-added full moon spectrum taken
for calibration purposes by CHIPS in narrow slit mode. 
The lines visible in the lunar spectrum arise from the solar corona 
being reflected by the moon.
The similarity in the spectral lines points
to an origin in the solar corona for the dayside lines in the 
non-lunar spectra as well.  
\label{moonspec}}
\end{figure}

Fig.~\ref{moonspec} shows the co-added lunar spectrum observed by 
CHIPS in narrow slit mode from several months in 2003, and 
the co-added spectrum from a number of observations when the
background lines were strongest.   The similarity of the spectra
is evident, although we would not expect identical spectra because
the solar spectrum itself varies and the scattered spectrum
was taken in wide-slit mode, which is more sensitive to the
scattered radiation and has lower spectral resolution. 
We can nonetheless use either the moon spectrum or the co-added
bright background times as a template to indicate 
the possible presence of solar-scattered lines in the following
way.  We developed this indicator by first subtracting the continuum 
level from the scattered light spectrum after 
fitting it with a polynomial.   We then similarly fit and subtract
the continuum from another CHIPS observation.    When the two
continuum-subtracted spectra are multiplied
together in the region of the lines, the template acts as a 
matched filter sensitive to
the solar lines.  We call the exposure time-normalized product of
the two spectra integrated over the spectral region 170 - 230 \AA\/ 
the "line flux indicator," or LFI.   We find LFI to be a sensitive
indicator of the presence of scattered solar light in co-added observations
and can use it to reject data that contains this background where
required.  The third trace in Fig.~\ref{moonspec} is co-added daytime data 
selected
for low LFI and shows essentially no solar contamination.  
However, in most cases we measure
only upper limits to the spectral lines expected from charge exchange 
reactions, so that the presence of a possible solar background is of less 
concern than it would be were CX lines detected. 
Additionally, the CX lines we are trying to detect form an only
partially overlapping set with the lines emerging from the solar
corona. 

\begin{figure}
\includegraphics[angle=90, scale=0.60]{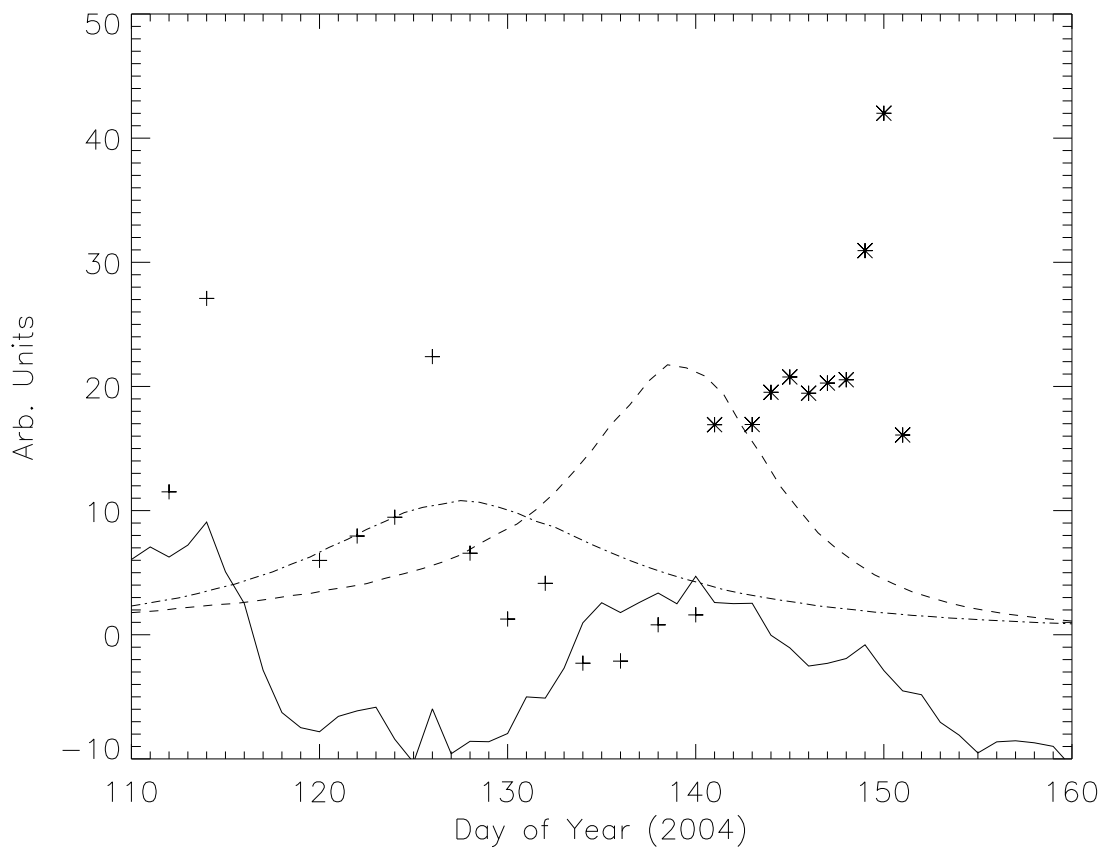}
\caption{The figure shows several quantities related to the faint, variable 
background versus day of year, 2004 in order to investigate 
correlations with the background.  The 
crosses indicate relative LFI (see text) strength by day for 
C/2001 Q4 (NEAT) observations and the asterisks for C/2002 T7 (LINEAR)
observations.  
The last six Q4 points are designated "no LFI" points.  The solid
line is a scaled and offset measure of the SEE 170 - 230 \AA\/ flux.  
The relative observing geometry appears to be a larger determinant
of LFI magnitude than the solar EUV flux as measured by SEE.
We also show, to indicate the relative comet brightness where our 
observations were made, a geometry-only prediction (based on Horizons
ephemeris) of comet flux based on distance from Earth and the Sun.   
The dot-dash line is for Q4 and the dashed line is for T7.  
\label{LFI}}
\end{figure}

We show in Fig.~\ref{LFI} the relative LFI value for each day of the
Q4 and T7 CHIPS observations.  We also show the scaled magnitude of
the solar 170 - 230 \AA\/ flux of the Sun as measured by SEE, 
which does not correlate with the LFI values.  Finally, we also
show a model visual brightness of the two comets, where the
variation is based only on the inverse-squared distances from the Earth 
to the comet and the comet to the Sun.  The latter curves indicate
that we observed Q4 through its peak brightness, while the
T7 observations were only possible as the brightness declined,
since the comet was too close to the Sun in the sky to observe prior 
to that time.  Owing to the sensitivity level of the CHIPS data, it
is not useful to subdivide the data into different time slices to 
investigate possible time variation of the emission.  

\subsection{$Chandra$ observations of C/2001 Q4 (NEAT)}

C/2001 Q4 (NEAT) was observed by $Chandra$ May 12, 2004 using director's 
discretionary
time.  The total approved observation of 10 ks was divided into 3 pointings
of approximately equal length.  During each observation the telescope was
held fixed and the comet allowed to drift through the field of view. The
comet was centered on the S3 chip which has the best low energy response.
The three resultant standard level 2 event lists were simultaneously
concatenated and reprojected into a fixed "comet-centered" frame of
reference using the CIAO\footnote{Chandra Interactive Analysis 
of Observations} tool "sso\_freeze".  The final reconstructed image
provides good spatial ($\sim$0.5"/pixel) and spectral information 
($\sigma\sim$50 eV).

A source spectrum was extracted from a circular aperture with radius of 
700 pixels (5.8$\arcmin$) centered on the peak of the comet
emission and the background 
spectra was extracted from the area outside the source aperture.  
We examined how varying the aperture radius changed the comet and 
background spectrum.  The aperture radius we chose included
essentially all of the emission from the comet; outside of this radius
we expect that heliospheric SWCX will be a uniform background on
this scale.    The background spectrum normalized by area 
was subtracted from the in-aperture spectrum to produce the Q4
comet spectrum.  Work by \citet{Lisse96} and \citet{Weg04} 
shows the CX X-ray emission from a comet as active as Q4 may be offset 
from the nucleus by 30,000 - 40,000 km.  At the distance of 0.6 au where 
Q4 was observed with $Chandra$, this separation translates to an angle 
of $\sim$1 arcminute.  Owing to the diffuse nature of the 
emission and the positional uncertainty induced by the photon 
remapping it is not clear whether we can discern a potential offset
of this magnitude from the center of the nucleus, but we have 
verified that the spectrum resulting from our chosen aperture 
radius is insensitive to a potential offset of this size.  
A detailed discussion of the Q4 X-ray emission morphology will be 
covered in \citet{Lisse07}.   ACIS response matrices to model
the instrument effective area and energy dependent sensitivity were created 
with the standard CIAO tools and the resulting spectra were fit using XSPEC.

\section{Results and Discussion}

Fig.~\ref{Q4summed} shows the comet C/2001 Q4 (NEAT) spectrum from 
50 - 270 \AA\/ measured with CHIPS.    
For each potentially strong charge exchange spectral line indicated in the 
figure, we measure the brightness
using the spectra from one or both halves 
of the CHIPS detector.  We have run our line flux measuring
algorithm for each comet and the results of these measurements are shown in
Table 2.  We include two measurements using Q4 data: the "no LFI"
set refers to the period May 9 - 19, 2004 corresponding to 
a time when the LFI measurement indicates minimal contamination from
scattered solar EUV radiation.  We also present the limits
from a spectrum consisting of the co-added spectra from all three 
comets, which contains the lowest relative shot noise background. 
We measure positive flux at the wavelengths of a number of
potential charge exchange lines, albeit at low significance.  

Fe lines at wavelengths 170 - 220 \AA\/
are detected in several of the spectra at significances 
ranging to 5 $\sigma$.  Since it is likely that much or all of
this emission is solar rather than cometary, we restrict 
our consideration to possible charge exchange lines shortward 
of 170 \AA\/, although line measurements that may have a solar 
component provide upper limits to the cometary emissions. 
We estimate that the lines for which we report limits 
in the table contribute about 90\% of the flux from charge 
exchange if we ignore the iron lines.  

The least contaminated spectrum is that obtained from the NEAT Q4
data selected for low LFI and the most significant line 
is the \ion{O}{6} line at 129.91 \AA\/, at almost 2$\sigma$, 
arising from the 1s4d $^3$D - 1s2p $^3$P transition. 

\begin{figure}
\includegraphics[angle=90, scale=0.60]{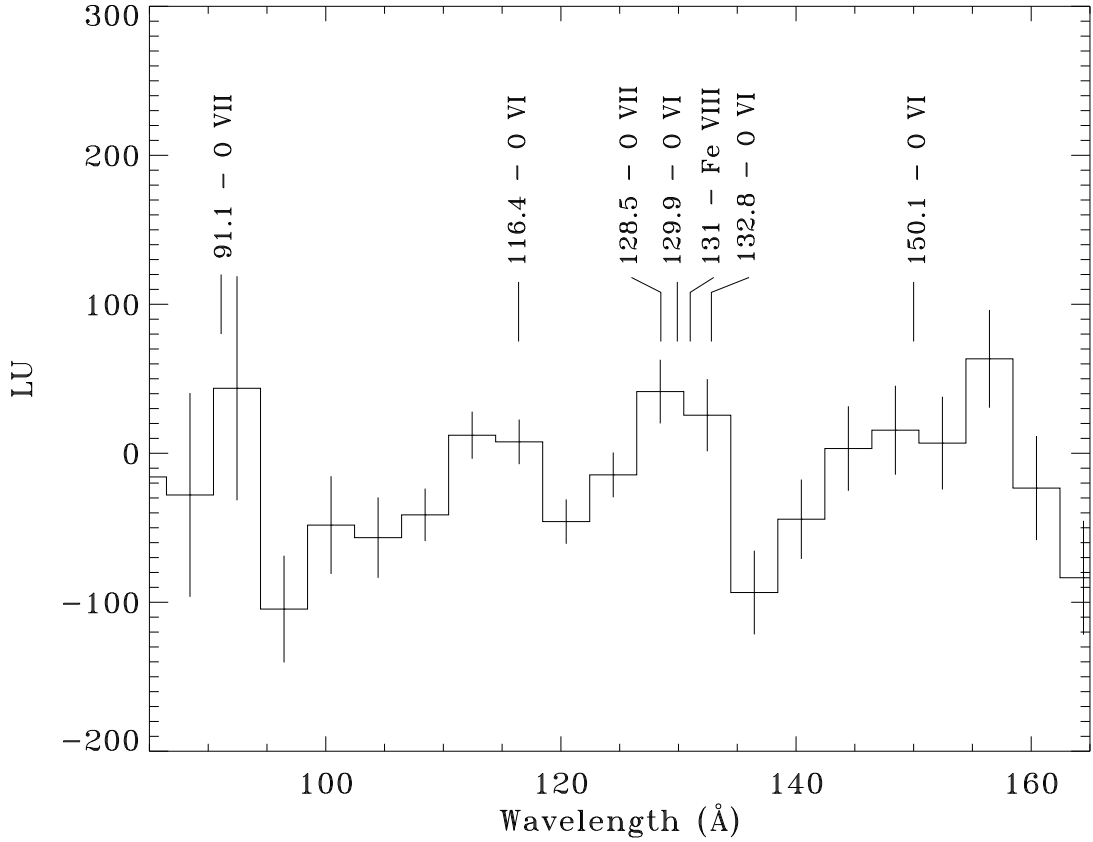}
\caption{All three comet spectra summed, then continuum subtracted.  This 
procedure can improve the visibility of strong emission lines, 
but is subject to uncertainties in the continuum level placement.  
Spectra and errors shown at 4 \AA\/ resolution, close to the actual 
instrument resolution and the error bars are 1 sigma.  
Some tentative spectral line identifications are indicated.  
Longward of 170 \AA\/, solar Fe lines dominate the spectrum. 
\label{allcomet}}
\end{figure}

In Fig.~\ref{allcomet} we show the results, binned to the
CHIPS instrumental resolution, of the line
flux measuring algorithm run on a 1 \AA\/ wavelength grid 
on the combined data set.  When using this technique a positive actual line
flux suppresses the neighboring continuum. Hence positive flux should be read
from the graph axis, not the adjacent continuum.
Tentative line identifications of features associated with 
positive flux measurements are indicated. 
The feature at 91.1 \AA\/ coincides with the transition 
1s4p $^3$P - 1s2s $^3$S of \ion{O}{7}, and the feature at 128.5 \AA\/ 
to the 1s3d $^3$D - 1s2p $^3$P transition. \ion{O}{6} has lines at 116.4 A,
129.9 \AA\/,132.5 \AA\/ and 150.1 \AA\/, corresponding respectively 
to the 4p-2s, 4d-2p, 4s-2p and 3p-2s transitions.

A search for line detections can also be conducted using
theoretical predictions for the wavelength and flux
ratio of several lines from the same ion. 
We employed the nine brightest lines in the CHIPS band of the \ion{O}{6} ion 
given by the model described in the following section.  
The line profiles were taken to be Gaussian with widths equal to 
the wavelength-dependent resolution of CHIPS.
The free parameters in the fit were a quintic polynomial to find the
background level and an overall amplitude for the \ion{O}{6} lines, with the 
individual line ratios fixed.  The results when this technique
was applied to the NEAT Q4 "no LFI" spectrum are that the line amplitudes
are consistent with zero. 
We carried out the same procedure for the co-added comet T7 spectra
and again the results were consistent with zero \ion{O}{6} emission. 

\section{Modeling the Solar Wind Charge Exchange Emission} \label{model}

\begin{figure}
\includegraphics[scale=0.90]{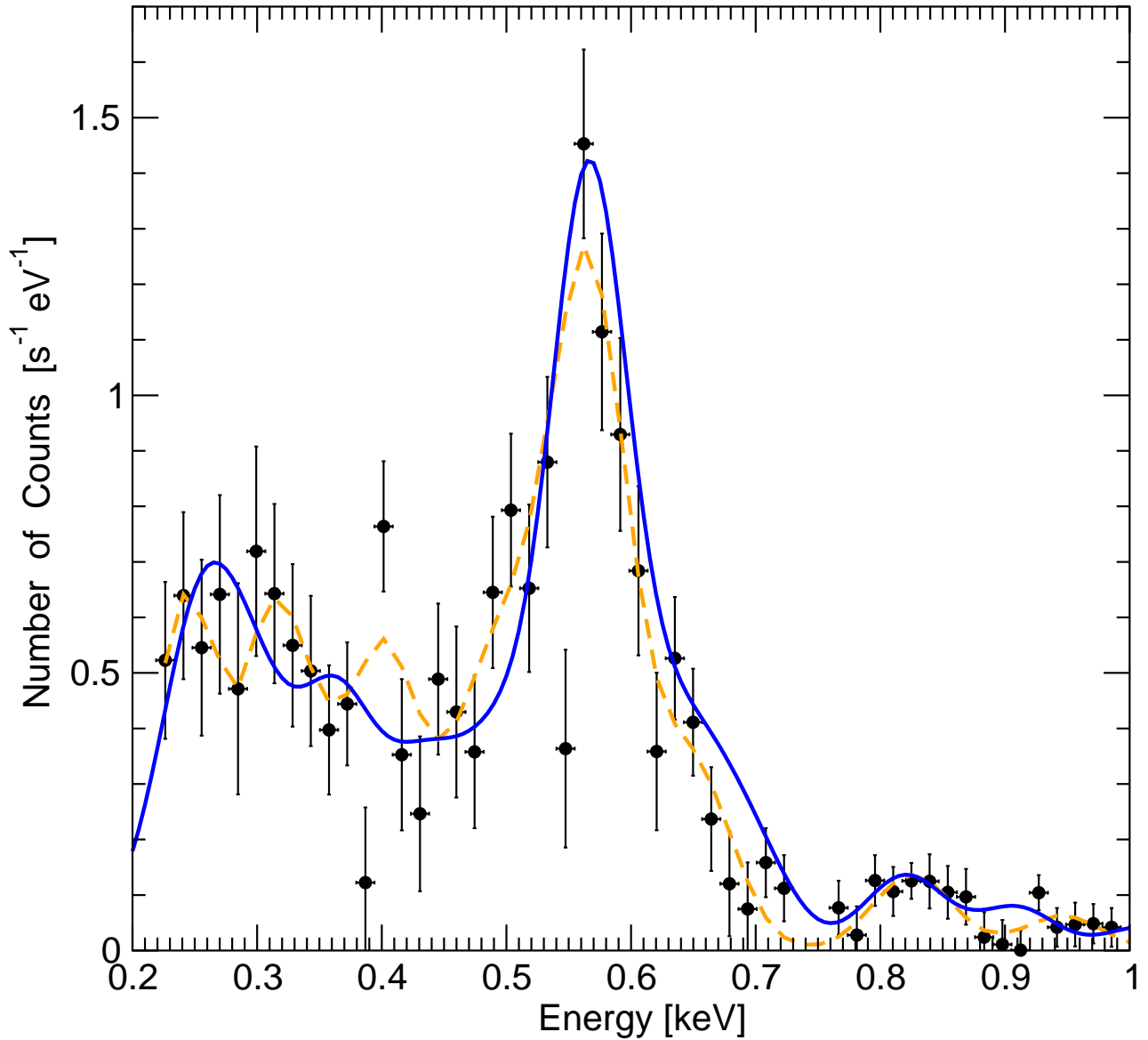}
\caption{Two models of X-ray emission lines are compared with
the $Chandra$ spectrum for C/2001 Q4 (NEAT).
The solid points are the number counts from $Chandra$ data.
The dashed curve is the $Chandra$ data interpreted
using an XSPEC fit with 8 lines between 245 eV and
952 eV folded through the instrument response.
The solid line shows the SWCX model predictions
at a resolution of 72 eV.  While both models
yield and acceptable fit from a $\chi^2$ standpoint,
only the SWCX model is based on substantially all of the
underlying physics of the emission process.
\label{swmodel}}
\end{figure}

The NEAT Q4 count spectrum obtained from CHANDRA is shown in Fig. 
~\ref{swmodel}. 
It contains emission lines seen in other comets \citep{Lisse01,
Kras02}.  We first use XSPEC to fit the
spectrum with eight emission lines whose wavelengths were allowed to vary.
The best fit was achieved with 8 lines at energies in eV of 245,
320, 403, 496, 570, 655, 830 and 952, with corresponding photon
fluxes in units of 10$^{-4}$ photons cm$^{-2}$ s$^{-1}$ of respectively
3.68, 26.5, 6.57, 2.59, 5.93, 1.05, 0.29, and 0.11.  
The resulting $\chi^2$ per degree of freedom was 0.91. The 
strong lines at 320, 403 and 570 eV correspond to transitions of 
\ion{C}{5}, \ion{N}{6}
and \ion{O}{7}.   The lines near 245 eV may be from \ion{Si}{9}, \ion{Mg}{9} 
or \ion{Mg}{10}, and a feature at 952 eV
found also in the spectra of Comets McNaught-Hartley
and Linear 4 \citep{Kras02, Kras04b}
may be from \ion{Ne}{9}.  There is an
indication of a line near 835 eV but we hesitate to attribute it to the
5p-1s resonance line of \ion{O}{8} because we do not find the 
corresponding 2p-1s line
at 653 eV which we predict should be a factor of six more intense. Because
of the low resolution of the instrument, fits with a similar 
$\chi^2$ figure of merit are also be obtained by a
combination of emission lines and thermal bremsstrahlung at a temperature
of about 0.23 keV, as in \citet{khar00}.

We have extended a model developed by \citet{khar00}, \citet{khar01}, 
\citet{khar03} and \citet{Pepino04} to compute emission 
spectra over the full EUV and X-ray spectral range. 
In the model we use electron capture cross sections for positive ions 
colliding
with neutral H$_2$O molecules that reproduce the experimental laboratory
X-ray spectra when available and theoretical estimates when they are
not \citep{Rigazio02}.  When applied to the measured spectra the method
takes into account empirically-determined processes such as multiple capture
followed by autoionization. The capture cross sections determine the
entry rates into the individual excited states of the ion created in the
charge exchange.  The excited states decay in a radiative cascade which
produces photons at X-ray, EUV and UV wavelengths.  The relative
intensities of the emission lines occurring in the cascade from a given
initial level depend only on the branching ratios of the transition
probabilities.  Thus the EUV spectrum is determined by the X-ray
spectrum and the comet spectra depend on and provide a measure of the solar
wind ion composition.  

\subsection{X-ray emission lines}

We calculated the X-ray emission line intensities for the ion
composition listed in Table ~\ref{ioncomp}.  It is based on the ion composition
of \citet{schwad00} for a fast solar wind,
supplemented by the presence of a fraction
0.001\% of highly charged Fe ions, Fe XV-Fe XX, that appear to be
necessary to reproduce the observed spectrum between 0.8 and 0.9 keV. 
An enhanced flux of highly-charged iron ions may be a
consequence of fast coronal mass ejection events \citep{Lepri04, 
Zur04}.

A comparison of the observations in counts per second of Comet NEAT
Q4 and the XSPEC and SWCX models at photon energies between 0.3 keV and 1.0 keV
is presented in Fig.5.  The CHANDRA data with the error bars indicating 
1 sigma uncertainties are shown as solid circles. The dashed line is the 
XSPEC model fit with 8 emission lines folded through the instrument response.   
The solid line is the SWCX prediction at a spectral resolution of 75 eV. 
In Table ~\ref{linesfit} we list the energies and wavelengths and the predicted relative 
intensities of the principal X-ray and EUV emission lines of the SWCX model. 
The agreement of the measurements of the X-ray lines with the models is 
good, but because of the low spectral resolution, we cannot exclude
a significant contribution from bremsstrahlung from XSPEC modeling of
the X-ray lines alone.

\subsection{EUV emission lines}

The photon fluxes measured for comet NEAT Q4 with CHIPS at energies
below 150 eV and with CHANDRA at energies between 250 eV and 1.0 keV  are
compared in Fig.6 with the predictions of the SWCX model. A scaling factor has
been applied to bring the predicted total flux of the X-ray emission
with energies greater than 250 eV into agreement with the measured value
of 3.7$\times 10^{-12}$ ergs cm$^{-2}$ s$^{-1}$.  Table ~\ref{linesfit} 
can then be 
used to yield the absolute intensities shown in Fig.6.  The corresponding 
total flux of the EUV lines between 47 eV and 140 eV is 
1.5$\times 10^{-12}$ ergs cm$^{-2}$ s$^{-1}$,
distributed amongst 181 lines of C,N,O, Ne and Fe. 
The theoretical EUV intensities fall well below
the measurements except near 95 eV, a region in which several lines of OVI
and OVII appear.  The most interesting CHIPS measurement of the SWCX 
lines is the 2 sigma detection in the  NEAT Q4 "no-LFI" spectrum 
of the line of OVI at 95.5 eV or 130 A.  Discrepancies at lower 
energies are affected by scattered radiation from the Sun.  We conclude 
that the SWCX model EUV predictions are consistent 
with the CHIPS data, but the instrument lacks the sensitivity needed to 
systematically test the theoretical predictions. 
In that CHIPS was not designed for measurements of comets, or
point sources, the limited agreement supports the SWCX model 
and points out some additional utility of the CHIPS measurements.
The model also predicts that the 10 brightest EUV lines in the 
CHIPS band contribute 65\% of all the EUV photons, and 88\% 
of the EUV photon flux is contained in the brightest 26 lines.  

The accumulated yield is useful in providing an estimate of the
photon energy distribution from comets. It is defined as the total number of
photons produced by charge exchange in collisions of the solar wind with
comets with energies lying between a minimum energy ,equal to 47 eV for
CHIPS, and a maximum energy E. The accumulated yield is dimensionless.
Multiplied by the product of the solar wind flux and ion density it yields the
volume emissivity.  Fig. 7 shows the calculated accumulated yield for comet NEAT
Q4 as a function of photon energy E for a minimum energy of 47 eV. 
Multiplying it by a factor of 26 yields the accumulated photon flux 
in units of cm$^{-2}$ s$^{-1}$.  The small values reflect the low fractional 
abundances of the heavy solar wind ions.

\section{Conclusions}

We have measured EUV emission from three comets and provided 
the first-ever simultaneous EUV and X-ray measurement and modeling
of cometary SWCX emission.  The X-ray data for comet Q4 show clear 
emission lines that are satisfactorily interpreted as 
solar wind charge exchange 
lines from ions expected in the solar wind. 
We have used the same model to identify and predict the strengths 
of EUV lines.  The measured EUV line strengths are consistent with the 
theoretical model that reproduces the X-ray line intensities. 
The CHIPS data and model results show that the EUV lines from 
all three comets are quite faint.  Nonetheless, the high spectral resolution
measurements of the EUV lines have now set useful limits on many
emission lines in this band that may serve as a guide for 
future EUV observations of comets. 


\acknowledgments

The CHIPS team gratefully acknowledges support by NASA Grant NAG 5-5213.
A.D. and V.K.  have been supported in this project by NASA grants NAG 5-1331
and NNG04GD57G.
We thank the SEE PI, team and NASA for making the SEE data available.

\clearpage 

\begin{deluxetable}{lllllllll}
\rotate
\tabletypesize{\scriptsize}
\tablecaption{Summary of CHIPS comet observations \label{tab:gen_obs}}
\tablewidth{0pt}
\tablehead{\colhead{Comet} &
\colhead{Date Range Obs.} &
\colhead{Exp. Time} &
\colhead{Exp. Time} &
\colhead{Sun Angle} &
\colhead{Sun-Comet} &
\colhead{Earth-Comet} &
\colhead{V mag. Range}   & 
\colhead{0.25-1 KeV Flux} \\
\colhead{} &
\colhead{} &
\colhead{Day (s)} &
\colhead{Night (s)} &
\colhead{Range (deg)} &
\colhead{Distance Range} &
\colhead{Distance Range} &
\colhead{} &
\colhead{ $10^{-12}$ ergs cm$^{-2}$ s$^{-1}$}\\
\colhead{} &
\colhead{(CHIPS)} &
\colhead{} &
\colhead{} &
\colhead{} &
\colhead{(AU)} &
\colhead{(AU)} &
\colhead{} &
\colhead{} } 
\startdata
C/2001 Q4 (NEAT) & 2004-04-21 07:35 - 2004-05-21 07:39 & 331096 & 5832  & 78.1 - 69.0 & 1.05 - 0.97 & 0.58 - 0.56 & 2.6 - 2.1 & 3.7 \\

C/2002 T7 (LINEAR) & 2004-05-02 11:45 - 2004-05-31 03:41 & 219477 & 14415 & 40.0 - 72.8 & 
0.65 - 1.0  & 0.76 - 0.55 & 1.5 - 2.7 & -  \\
C/2004 Q2 (Machholz) & 2005-01-06 23:44 - 2005-01-28 22:58 & 441466 & 41158 & 130.3 - 106.7 
& 1.24 - 1.21 & 0.35 - 0.47 & 4.1 - 4.7 & - \\
\enddata
{\flushleft Column 9 from $Chandra$ observations of May 9, 2004}
\end{deluxetable}

\clearpage

\begin{deluxetable}{lllllllllllllllll}
\tabletypesize{\scriptsize}
\rotate
\tablewidth{0pt}
\tablecolumns{14}
\tablecaption{Charge Exchange Line Flux Limits \label{tab:line_limits}}
\tablehead{
\colhead{}  &  &  \multicolumn{3}{c}{C/2001 Q4 no LFI} & \multicolumn{3}{c}{C/2001 Q4} & 
\multicolumn{3}{c}{C/2002 T7}  & \multicolumn{3}{c}{C/2004 Q2} & 
\multicolumn{3}{c}{All Combined} \\
\cline{3-5} \cline{6-8} \cline{9-11} \cline{12-14} \cline{15-17} \\
\colhead{Wavelength (\AA)} &
\colhead{Ion} &
\colhead{Limit} &
\colhead{$\sigma$ } &
\colhead{Sign.} &
\colhead{Limit } &
\colhead{$\sigma$ } &
\colhead{Sign.} &
\colhead{Limit } &
\colhead{$\sigma$ } &
\colhead{Sign.} &
\colhead{Limit } &
\colhead{$\sigma$ } &
\colhead{Sign.} &
\colhead{Limit } &
\colhead{$\sigma$ } &
\colhead{Sign.} } 
\startdata
98.37 &	Ne {\sc VIII} & -0.22 & 1.63 & 0.00	 & -0.87 & 1.03 & 0.00	 & -2.58 & 1.23 & 0.00	 & -1.00 & 0.89 & 0.00	 & -1.30 & 0.63 & 0.00 \\
115.86 & O {\sc VI} & -0.51 & 0.77 & 0.00	 & 0.26 & 0.48 & 0.53	 & 1.01 & 0.58 & 1.74	 & 0.19 & 0.42 & 0.45	 & 0.39 & 0.30 & 1.31 \\
120.33 & O {\sc VII} & 0.20 & 0.76 & 0.27	 & 0.14 & 0.47 & 0.29	 & -0.83 & 0.56 & 0.00	 & -0.62 & 0.41 & 0.00	 & -0.42 & 0.29 & 0.00 \\
128.48 &O {\sc VII} & 0.26 & 1.07 & 0.25	 & 0.06 & 0.66 & 0.09	 & 0.10 & 0.78 & 0.12	 & 0.64 & 0.56 & 1.13	 & 0.33 & 0.40 & 0.84 \\
129.91\dag &O {\sc VI} & 2.22 & 1.16 & 1.92	 & 0.52 & 0.71 & 0.73	 & -0.34 & 0.83 & 0.00	 & 0.56 & 0.60 & 0.93	 & 0.35 & 0.43 & 0.82 \\
132.35\dag &O {\sc VI} & 0.56 & 1.24 & 0.45	 & 0.83 & 0.76 & 1.10	 & 0.14 & 0.89 & 0.15	 & -0.29 & 0.64 & 0.00	 & 0.17 & 0.46 & 0.36 \\
133.92 &N {\sc VII} & -0.04 & 1.29 & 0.00	 & 1.07 & 0.80 & 1.34	 & 0.40 & 0.94 & 0.42	 & -0.69 & 0.67 & 0.00	 & 0.11 & 0.48 & 0.24 \\
135.02 &C {\sc VI} & -0.27 & 1.31 & 0.00	 & 0.27 & 0.81 & 0.34	 & 0.34 & 0.95 & 0.36	 & -0.08 & 0.68 & 0.00	 & 0.12 & 0.48 & 0.26 \\
150.21 &O {\sc VI} & -1.84 & 1.60 & 0.00	 & -0.21 & 0.97 & 0.00	 & -1.09 & 1.14 & 0.00	 & 0.15 & 0.82 & 0.19	 & -0.24 & 0.58 & 0.00 \\
173.29 &O {\sc VI} & 1.84 & 2.02 & 0.91	 & 5.16 & 1.26 & 4.10	 & 3.48 & 1.57 & 2.22	 & 1.85 & 0.87 & 2.13	 & 3.27 & 0.73 & 4.49 \\
182.28 &C {\sc VI} & 0.64 & 2.56 & 0.25	 & 2.76 & 1.63 & 1.69	 & 6.13 & 2.11 & 2.91	 & -0.69 & 1.02 & 0.00	 & 1.90 & 0.95 & 2.00 \\
184.19 &O {\sc VI} & -2.95 & 2.56 & 0.00	 & -1.96 & 1.63 & 0.00	 & -7.77 & 2.10 & 0.00	 & 0.55 & 1.03 & 0.54	 & -2.06 & 0.95 & 0.00 \\
186.74 &C {\sc V} & 2.15 & 2.60 & 0.83	 & -0.47 & 1.65 & 0.00	 & -0.47 & 2.14 & 0.00	 & 1.59 & 1.04 & 1.53	 & 0.48 & 0.97 & 0.50 \\
197.05 &C {\sc V} & 0.11 & 2.80 & 0.04	 & 6.09 & 1.80 & 3.37	 & 11.65 & 2.36 & 4.93	 & 0.50 & 1.12 & 0.45	 & 4.72 & 1.05 & 4.48 \\
209.44 &N {\sc V} & 2.15 & 3.20 & 0.67	 & -2.88 & 2.03 & 0.00	 & -3.61 & 2.61 & 0.00	 & 1.81 & 1.27 & 1.43	 & -0.87 & 1.18 & 0.00 \\
227.20 &C {\sc V} & 4.84 & 3.14 & 1.54	 & 3.42 & 1.96 & 1.75	 & 5.15 & 2.43 & 2.12	 & -1.01 & 1.23 & 0.00	 & 1.75 & 1.13 & 1.55 \\
248.75 &C {\sc V} & 8.32 & 4.38 & 1.90	 & 2.06 & 2.74 & 0.75	 & 7.36 & 3.40 & 2.16	 & -6.14 & 1.76 & 0.00	 & -0.57 & 1.60 & 0.00 \\
\enddata
{\flushleft{Table 2 notes:
Limit and local $\sigma$ values in LU. Sign. = Limit/$\sigma$. To find a limit in
ph cm$^{-2}$ s$^{-1}$ multiply the value in LU by 0.0106.
\newline
\dag Indicated lines and those with $\lambda >$ 170 \AA\/ are subject to contamination from solar Fe VIII - Fe XII emission }}
\end{deluxetable}

\clearpage
\begin{deluxetable}{l|l|l|l|l|l|l|l|l|l}
\tablecolumns{2}
\tablewidth{0pt}
\tablenum{3}
\tablecaption{Solar Wind Ion Composition \label{ioncomp}}
\rotate
\tablehead{
\colhead{Atom} & \multicolumn{9}{c}{Ionization State} \\
\colhead{} & \colhead{5+} 
& \colhead{6+} 
& \colhead{7+}
& \colhead{8+}
& \colhead{9+}
& \colhead{10+}
& \colhead{11+}
& \colhead{12+}
& \colhead{13+ - 20+} \\
}
\startdata
C  & 0.058 & 0.004 & & & & & & & \\
N  &0.0074 & 0.0012  & 0.0009 & & & & & &  \\
O &  & 0.053  & 0.004 & & & & & &  \\
Ne & & &  & 0.006&  0.00006 & & & & \\
Mg & & & & & 0.0025  &  0.0017 & & & \\ 
Si & & &  & 0.0015 & 0.0014 & 0.00007 & & & \\ 
S & & & & & 0.0004 & & & & \\
Fe & & & & & 0.0015 &  0.015 & 0.0001 & 0.0003 & 0.001  \\
\enddata
{\flushleft{Table 3 notes: Fractional abundance by number density
for each ion used for X-ray and EUV model spectrum.  
Blank columns indicate zero abundance was used for the model spectrum. }}
\end{deluxetable}

\begin{deluxetable}{llll}
\tablecolumns{4}
\tablewidth{0pt}
\tablenum{4}
\tablecaption{Relative Intensities of Emission Lines \label{linesfit}} 
\tablehead{
\colhead{Energy (eV)}
& \colhead{Rel. Intensity}
& \colhead{Ion}
& \colhead{Transition} }
\startdata
47.65    &  0.19  & \ion{C}{5}  & 1s3s $^3$S - 1s2p $^3$P \\
49.85    &  0.34  & \ion{C}{5}  & 1s3d $^3$D - 1s2p $^3$P \\
54.58    &  0.30  & \ion{C}{5}  & 1s3p $^3$P - 1s2s $^3$S \\
71.56    &  0.50  & \ion{O}{6} & 3d - 2p \\
82.56    &  0.20  & \ion{O}{6} & 3p - 2s \\
93.70    &  0.31  & \ion{O}{6} & 4s - 2p \\
95.43    &  0.19  & \ion{O}{6} & 4d - 2p \\
107.03   &  0.41  & \ion{O}{6} & 4p - 2s \\
298.96   &  1\tablenotemark{a} & \ion{C}{5} & 1s2s $^3$S - 1s$^2$ $^1$S \\
304.25   &  0.07  & \ion{C}{5}  & 1s2p $^3$P - 1s$^2$ $^1$S \\
307.91   &  0.16  & \ion{C}{5}  & 1s2p $^1$P - 1s$^2$ $^1$S \\
354.52   &  0.07  & \ion{C}{5}  & 1s3p $^1$P - 1s$^2$ $^1$S \\
367.35   &  0.11  & \ion{C}{6} & 2p - 1s \\
370.92   &  0.03  & \ion{C}{5}  & 1s4p $^1$P - 1s$^2$ $^1$S \\
419.80   &  0.03  & \ion{N}{6} & 1s2s $^3$S - 1s$^2$ $^1$S \\
435.38   &  0.03  & \ion{C}{6} & 3p - 1s \\
500.01   &  0.03 & \ion{N}{7} & 2p - 1s \\
560.86   &  0.14 & \ion{O}{7} & 1s2s $^3$S - 1s$^2$ $^1$S \\
\tablenotetext{a} {The intensity  of the \ion{C}{5} line  corresponds to a photon flux
of 3.3$\times 10^{-3}$ photons cm$^{-2}$s$^{-1}$.}
\enddata
{\flushleft{Table 4 lists the transitions with the highest emission
by photon number in the model spectrum.
The electronic configurations shown in the table for Li-like and H-like
ions represent transitions for all states of electronic doublets.} }
\end{deluxetable}

\clearpage


\begin{figure}
\includegraphics[scale=0.90]{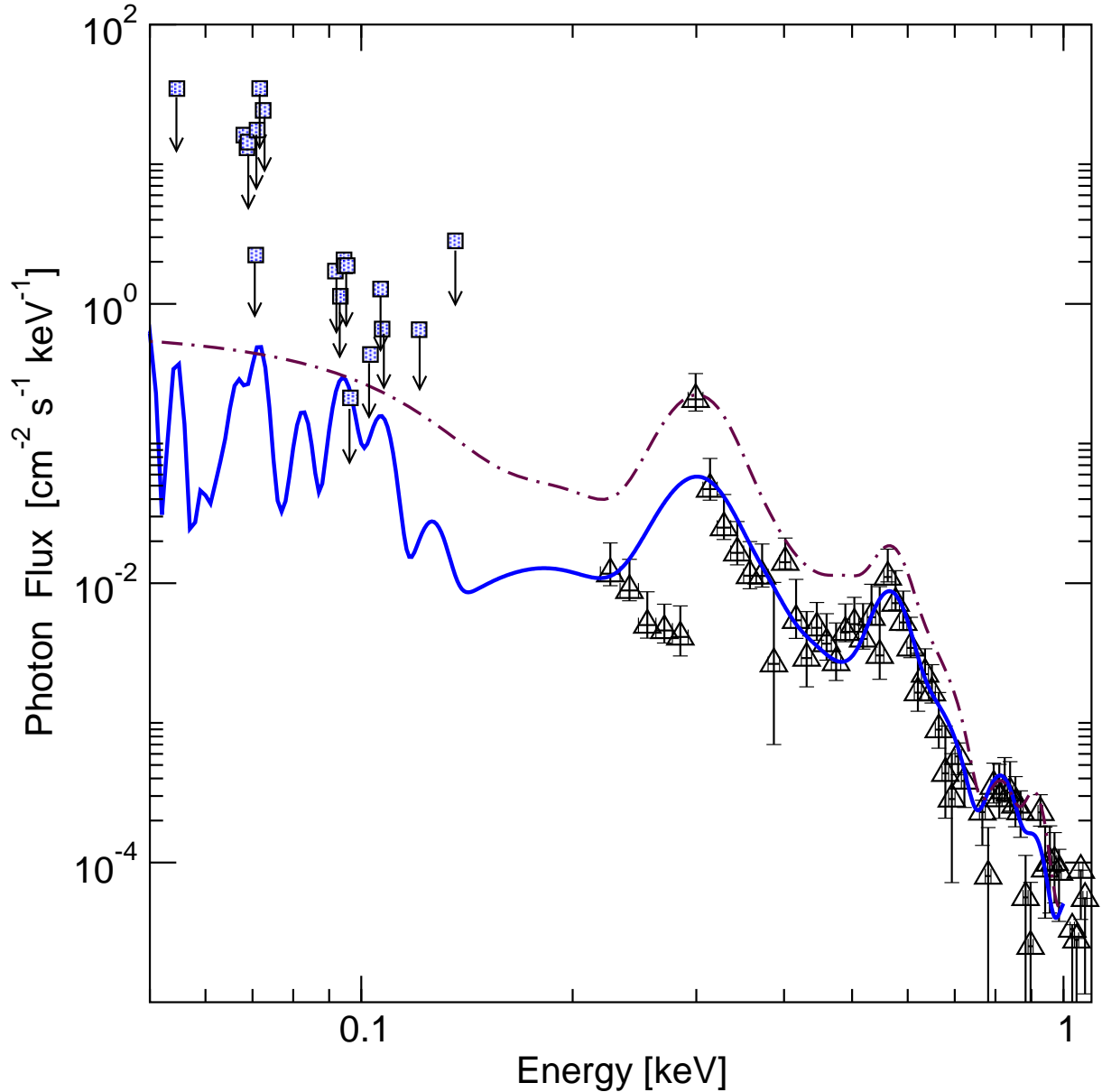}
\caption{The intensity of the EUV and X-ray emission induced
by the solar wind ions interacting with the atmosphere of 
Comet C/2001 Q4 (NEAT).
The solid curve shows the prediction of the CX model for
the photon flux in the energy interval of 0.047 - 1.1 keV.  
Absolute intensities of the theoretical CX spectra are
normalized to the total number of X-ray photons 
observed with the $Chandra$ telescope in the energy
interval 0.25 - 1 keV.  The triangles with
error bars represent $Chandra$ observations and the squares 
show the result of observations with CHIPS. 
Theoretical X-ray spectra are computed for a FWHM
of 70 eV, which corresponds to the Chandra observations.  
The resolution of the EUV spectral lines between 47 and 138 eV
has been taken to be equal to the CHIPS resolution of 4 \AA\/. 
The dot-dash line is the model fit to the unsubtracted comet
X-ray spectrum including both comet and heliospheric charge
exchange components; it shows a predicted absolute upper limit 
to the SWCX emission.  The EUV part of this curve is displayed
with the same resolution as the X-ray portion. 
\label{euvmodeldata}}
\end{figure}

\begin{figure}
\includegraphics[scale=0.95]{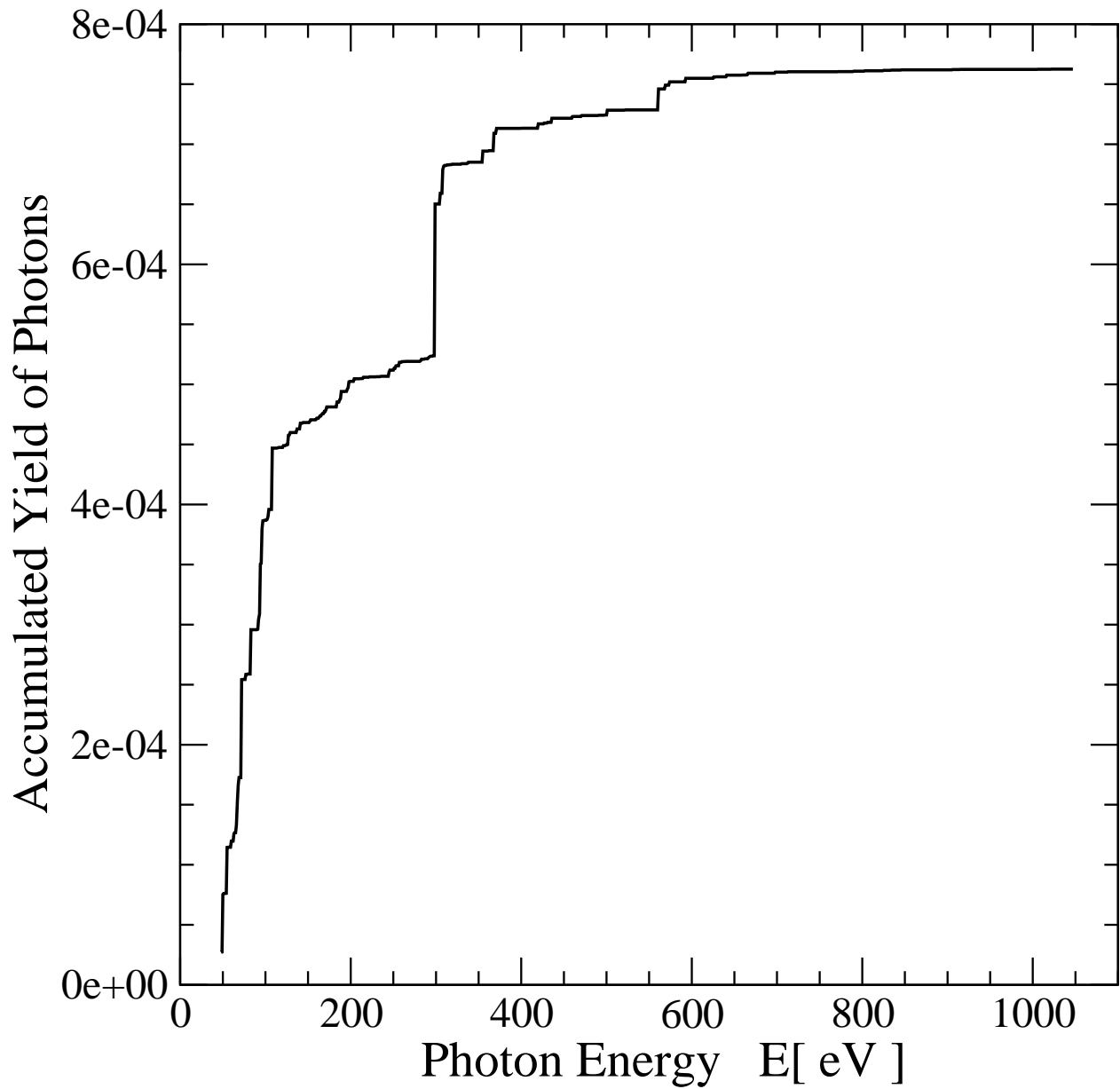}
\caption{The predicted Accumulated Yield of 
EUV and X-ray photons for a single charge exchange
collision of the heavy solar wind ions
calculated from the SWCX model.  The figure shows the
average number of photons induced with energies between 47 eV and
E[eV] in a single CX collision of the solar wind ions with molecules of a
cometary gas.  The averaging procedure takes into account the relative
abundances  of the solar wind protons, alpha particles and heavy ions,
and individual CX cross sections.  
\label{accumyield}}
\end{figure}

\end{document}